\documentclass[aps,prl,twocolumn,superscriptaddress]{revtex4-1}
\usepackage{mathrsfs}
\usepackage{epsfig}
\usepackage{graphicx}
\usepackage{amsfonts}
\usepackage[figuresright]{rotating}
\usepackage{amssymb}
\usepackage{amsmath}
\usepackage{dcolumn}
\usepackage{bm}
\usepackage{color}

\def\Im{\rm{Im}}
\def\be{\begin{equation}} \def\ee{\end{equation}}
\def\bea{\begin{eqnarray}} \def\eea{\end{eqnarray}}

\def\clb{\color{black}}

\newcommand{\sjtu} {Key Laboratory of Artificial Structures and Quantum
Control, Department of Physics and Astronomy, Shanghai Jiao Tong University, Shanghai 200240, People's Republic of China}

\newcommand{\zhiyuan} {Zhiyuan College, Shanghai Jiao Tong University, Shanghai 200240, China}

\newcommand{\WQCASQC} {Wilczek Quantum Center, School of Physics and Astronomy, Shanghai Jiao Tong University, Shanghai 200240, China}

\newcommand{\SRCQC}{Shanghai Research Center for Quantum Sciences, Shanghai 201315, China}

\begin{document}
\title{Pattern formation and exotic order in driven-dissipative Bose-Hubbard systems}

\author{Zijian Wang}
\thanks{These authors contributed equally to this work.}
\affiliation{\WQCASQC}
\affiliation{\zhiyuan}

\author{Carlos Navarrete-Benlloch}
\thanks{These authors contributed equally to this work.}
\affiliation{\WQCASQC}
\affiliation{\SRCQC}

\author{Zi Cai}
\email{zcai@sjtu.edu.cn}
\affiliation{\WQCASQC}
\affiliation{\SRCQC}
\affiliation{\sjtu}

\begin{abstract}
 Modern experimental platforms such as supercoducting-circuit arrays call for the exploration of bosonic tight-binding models in unconventional situations with no counterpart in real materials. Here we investigate one of such situations, in which excitations are driven and damped by pairs, leading to pattern formation and exotic bosonic states emerged from a non-equilibrium quantum many-body system. Focusing on a two-dimensional driven-dissipative Bose-Hubbard model, we find that its steady states are characterized by the condensation of bosons around momenta lying on a ``Bose surface'', a bosonic analogue of the Fermi surface in solid-state systems. The interplay between instabilities generated by the driving, the nonlinear dissipative mode-coupling, and the underlaying lattice effect, allows the system to equilibrate into an exotic superfluid state of bosons condensed on a closed ring in momentum space instead of discrete points.  Such an unconventional state with a spatially  uniform density distribution goes beyond the traditional scope of pattern formation, and thus has no counterpart in the classical literature. In addition, it is a state connected to several open problems in modern condensed-matter physics, and here we provide the means to stabilize it, opening the way to its experimental study. Moreover, we also provide a concrete experimental implementation of our model in currently-available superconducting-circuit arrays. We also investigate the relaxation spectrum around the condensate, which shows a characteristic purely diffusive behavior.
\end{abstract}


\maketitle

{\it Introduction}.---The scope of non-equilibrium physics is immense since the universe as a whole is a non-equilibrium system. A fundamental question in this context is understanding how the observed richness of spatiotemporal patterns spontaneously emerges from nothing \cite{Cross2009}. In contrast to pattern formation within thermodynamic equilibrium, rooted in the minimization of (free) energy, patterns emerging in non-equilibrium systems can only be understood within a dynamical framework,  even if the patterns of interest are time-independent. More often than not, when a system is driven far from equilibrium, spatially-uniform structures become unstable toward the growth of small perturbations, which leads to dynamics that amplify fluctuations and increase complexity. Late-time dynamics is dominated by the fastest-growing fluctuating modes, whose characteristic length and time scales determine the resulting spatiotemporal patterns, eventually stabilized by nonlinear and dissipative mechanisms \cite{Cross1993}. In such a dynamical framework, dynamical instabilities and nonlinear mode coupling mechanisms are crucial for pattern formation \cite{Swift1997}.

Nonequilibrium pattern formation has been intensively studied in classical systems ranging from hydrodynamics \cite{Drazin2004} and cosmology \cite{Liddle2000}, to biochemistry \cite{Koch1994} and optics \cite{Staliunas03book,Mandel97book,Arecchi91,Weiss07}. A profound question is then how to generalize these ideas to non-equilibrium quantum systems, where the interplay between the intrinsic quantum fluctuations and external non-equilibrium conditions might give rise to richer phenomena than what is expected on the basis of these effects separately \cite{PerezArjona06,PerezArjona07,Navarrete08}. The situation is further complicated and potentially richer when the quantum system is an interacting many-body system, opening avenues for observing exotic quantum states of matter that are absent in either its equilibrium quantum counterparts or in non-equilibrium classical systems. Recently, significant experimental progress has been made in Bose-Einstein condenstates (BECs), where stripes, squares, hexagons, and other types of patterns have been observed in exciton polaritons \cite{Ardizzone2013,Ma17} and ultracold atoms \cite{Kronjager2010,Hung2013,Kadau2016,Cosme2018,zhang2019,Vidmar2015,Clark2016,Sheikhan2019,Leonard17}.  But besides these conventional patterns, it is even more interesting to investigate exotic non-equilibrium states inspired by the intrinsic quantum nature of these systems, that have not been discussed in their classical counterparts.

In this work, we study pattern formation and exotic order in the non-equilibrium steady states of a pair-driven-dissipative Bose-Hubbard (BH) model, for which we propose a concrete implementation based on current superconducting-circuit arrays. In contrast to the continuous systems studied previously \cite{Ardizzone2013,Ma17}, here we investigate a tight-binding model defined on a two-dimensional (2D)  square lattice, where many-body effects are known to play a crucial role in determining equilibrium phase diagrams \cite{Fisher1989}. To drive the system out of equilibrium, we consider local pair creation/annihilation terms (pair driving), which induce spatially dependent instabilities determined by the fastest growing modes, which we show to lay on the Bosonic analogue of a Fermi surface. We include the nonlinear dissipation that unavoidably accompanies pair driving, and serves to stabilize the system. We consider two distinct situations. First, that in which the Bose surface is a generic closed curve, leading to unconventional superfluid states forming striped density patterns. Then, we consider a so-called nested surface, for which we obtained an exotic state with bosons condensed on a closed ring instead of discrete points in the Brillouin zone, leading to a spatially uniform density, but with a nontrivial phase distribution. In equilibrium physics, similar Bose-liquid states have been conjectured to play an important role in  frustrated quantum magnetism \cite{Sedrakyan2015}, high-$T_c$ superconductors \cite{Jiang2019}, and cold atoms with spin-orbit coupling \cite{Wu2011,Gopalakrishnan2011}. We also discuss the relaxation spectrum of fluctuations around the generic condensate, showing that it is dominated by a purely diffusive mode.


{\it Model and method}.---We study a 2D BH model in a square lattice with on-site pair creation/annihilation, governed by the Hamiltonian

\begin{equation}
{\clb\frac{\hat{H}}{\hbar}}=-J\sum_{\langle \mathbf{ij}\rangle}\hat{b}^\dag_\mathbf{i} \hat{b}_\mathbf{j}+\frac12\sum_\mathbf{i}\left(\frac U2 \hat{b}_\mathbf{i}^{\dagger 2}\hat{b}_\mathbf{i}^2-\nu\hat{n}_\mathbf{i}+\frac\Delta2\hat{b}_\mathbf{i}^2\right)+\text{H.c.},\label{eq:ham}\small
\end{equation}
where $\hat{b}_\mathbf{i}$ annihilates a boson at site $\mathbf{i}$ and $\hat{n}_\mathbf{i}=\hat{b}_\mathbf{i}^\dag \hat{b}_\mathbf{i}$ is the corresponding number operator. $J$ is the single-particle hopping rate between adjacent lattice sites $\langle\mathbf{ij}\rangle$.  $\nu$ resembles the chemical potential of equilibrium systems, but in our nonequilibrium setup it can be tuned from positive to negative \cite{SupMat}. $\Delta$ is the pair-driving rate, which we take positive without loss of generality. In a conventional BH model, $U$ is the interaction rate, but in our dissipative model it will adopt a more general meaning that we discuss later.

To get a better understanding of the effect of pair driving, we focus first on the $U=0$ case. The Hamiltonian takes a quadratic form with translational invariance, which is written in momentum space as
\begin{equation}
{\clb\frac{\hat{H}}{\hbar}}=\sum_\mathbf{k} (\varepsilon_\mathbf{k}-\nu) \hat{b}_\mathbf{k}^\dag \hat{b}_\mathbf{k}+\frac\Delta2 (\hat{b}_\mathbf{k}^\dag \hat{b}_{-\mathbf{k}}^\dag+\hat{b}_\mathbf{k}\hat{b}_{-\mathbf{k}})\label{eq:hamk}
\end{equation}
where the sum extends over momenta in the first Brillouin zone and $\hat{b}_\mathbf{k}=\frac 1L\sum_{\mathbf{i}}e^{-i\mathbf{k}\cdot \mathbf{i}}\hat{b}_\mathbf{i}$, for an $L\times L$ lattice with dispersion $\varepsilon_\mathbf{k}=-2J(\cos k_x+\cos k_y)$. Eq. (\ref{eq:hamk}) shows that each pair of $\pm\mathbf{k}$-modes with opposite momentum evolves independently with a Hamiltonian reminiscent to that of a detuned parametric amplifier \cite{Mollow1967,Carmichael84}. The corresponding physics is easily understood by analyzing the amplitudes $\psi_\mathbf{k}=\langle \hat{b}_\mathbf{k}\rangle$, with equations of motion
\begin{equation}
 i\frac{d}{dt}\left(\begin{array}{c}
    \psi_\mathbf{k}   \\
    \psi^*_{-\mathbf{k}} \\
\end{array}\right)=  \left(
  \begin{array}{cc}
    \varepsilon_\mathbf{k}-\nu & \Delta   \\
  -\Delta &\nu-\varepsilon_\mathbf{k}
\end{array}
\right)
\left(
  \begin{array}{c}
    \psi_\mathbf{k}   \\
    \psi^*_{-\mathbf{k}} \\
\end{array}
\right),\label{eq:EOM}
\end{equation}
Their general solution can be written as $\psi_\mathbf{k}(t)=e^{i\lambda_\mathbf{k}t}u_\mathbf{k}+e^{-i\lambda_\mathbf{k}t}v_\mathbf{k}$, where $u_\mathbf{k}$ and $v_\mathbf{k}$ are time-independent coefficients determined by the initial conditions and $\lambda_\mathbf{k}^2=(\varepsilon_\mathbf{k}-\nu)^2-\Delta^2$. Those $\mathbf{k}$-modes satisfying $|\varepsilon_\mathbf{k}-\nu|\geq \Delta$ evolve in a stable  fashion. In contrast, the modes with $|\varepsilon_\mathbf{k}-\nu|< \Delta$ are dynamically unstable and diverge exponentially with time. The divergence rate $\Im\{\lambda_\mathbf{k}\}$ is maximized for the $\mathbf{k}$-modes satisfying $\varepsilon_\mathbf{k}-\nu=0$, which for fermionic models corresponds to the Fermi surface, and we thus dub ``Bose surface'' here.

The instability at $U=0$ indicates that the density of bosons will increase indefinitely.  In a real system, however, dissipation and nonlinear effects (interactions) make the density saturate, eventually halting the system into a steady state. In particular, the pair driving that we consider here will be accompanied by  two-boson (nonlinear) loss in real implementations, as we highlight in \cite{SupMat}. Mathematically, this has to be treated through a master equation for the mixed state of the system.  However, under the assumption that superfluid order is present, we can simplify the problem by invoking the mean-field or coherent-state approximation. As detailed in \cite{SupMat}, on the one hand this is equivalent to adding an imaginary part to the interaction, that is, $U=g-i\gamma$ with $g$ and $\gamma$ real and positive, which makes the Hamiltonian (\ref{eq:ham}) non-Hermitian, becoming then an effective description of the open system.  On the other hand, the coherent-state approximation amounts to replacing the bosonic operators by their expectation value $\psi_\mathbf{i}=\langle \hat{b}_\mathbf{i}\rangle$ in the Heisenberg equations. Since our model is defined on a lattice, this leads to a finite-differences version of the Gross-Pitaevskii (GP) equation:
\begin{equation}
i\frac{d\psi_\mathbf{i}}{dt}=-J\sum_\mathbf{j}\psi_\mathbf{j}-\nu\psi_\mathbf{i}+(g-i\gamma)|\psi_\mathbf{i}|^2\psi_\mathbf{i}+\Delta \psi_\mathbf{i}^*, \label{eq:GP}
\end{equation}
where the summation is restricted to the sites $\mathbf{j}$ adjacent to site $\mathbf{i}$. More often than not, driving and dissipation inevitably heat up the system, and are thus detrimental to superfluid order. However, focusing on the thermodynamic limit with infinite boson numbers (where dissipative tunneling between symmetry-breaking states takes an infinite time \cite{Kinsler91,Navarrete17cycles,Iemini18}), and a regime where the driving, dissipation, and interaction rates are much smaller than the hopping rate ($\Delta,g,\gamma\ll J$), superfluidity is expected to survive in the non-equilibrium steady state. Indeed, this is supported by experimental observations in exciton-polariton BECs \cite{Carusotto2013,Ardizzone2013,Ma17} and theoretical analysis based on complex GP equations \cite{Wouters2007,Tauber2014}.

Note as well that for weakly-interacting bosonic models, it is known that lattice effect is not important for ground states,  which are usually superfluid states with bosons condensed at zero momentum, irrespective of the lattice geometry. In contrast, we show below that the lattice effect plays an important role in our non-equilibrium steady state, particularly through the Bose-surface nesting effect, which is absent in continuous space or non-bipartite lattice (e.g. triangle lattice).

\begin{figure*}[htb]
\includegraphics[width=0.99\linewidth]{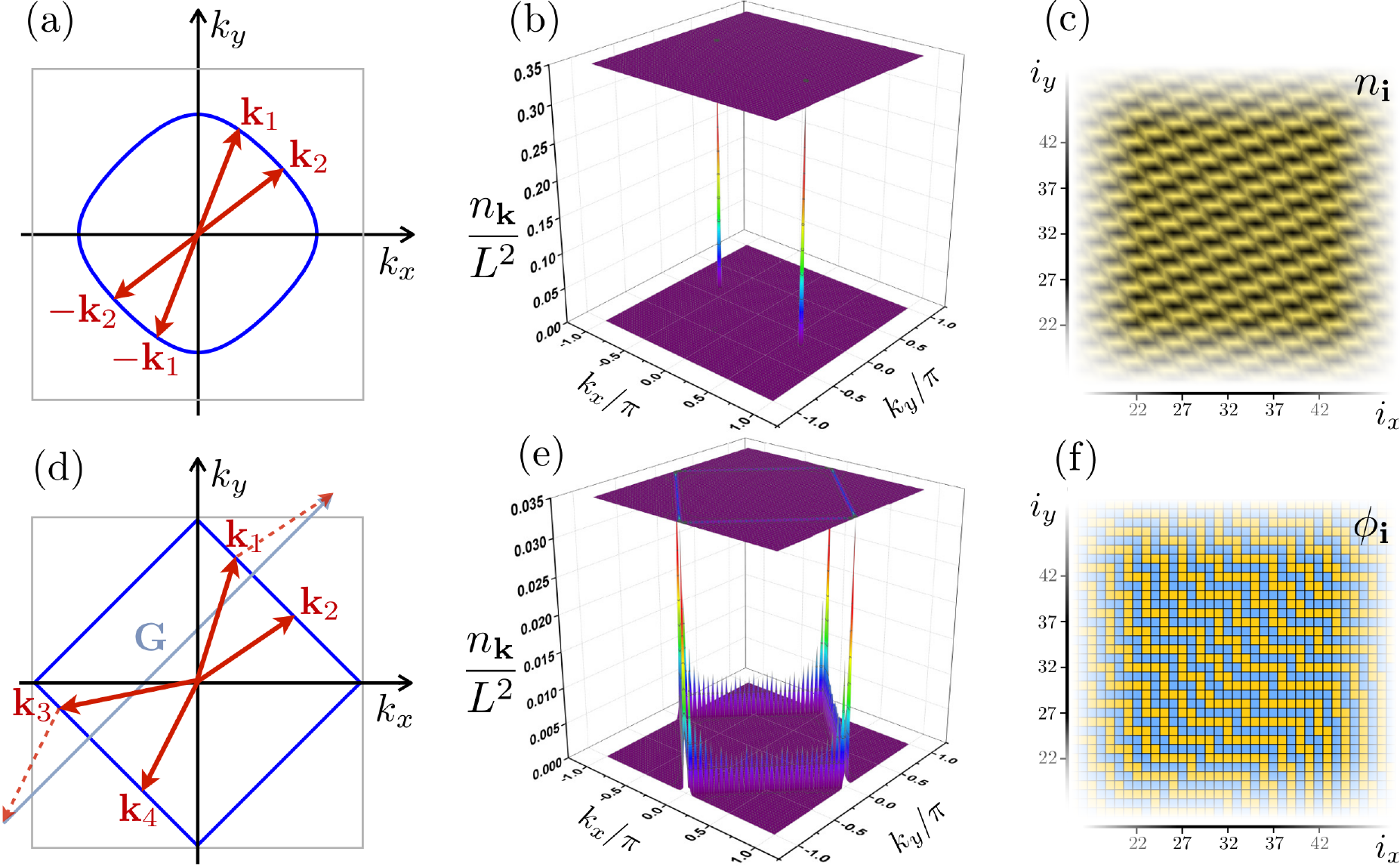}
\caption{(Color online) Two characteristic Bose surfaces (blue solid line, determined by $\varepsilon_\mathbf{k}-\nu=0$) with (a) generic geometry with $\nu\neq 0$ and (d) nested geometry with $\nu=0$, whose opposite contours are connected by a $\mathbf{G}/2=(\pi,\pi)$ vector. In (b) and (c) we show the steady-state density distribution in momentum and real space (lighter colors correspond to larger densities), respectively, for a generic geometry with $\nu=-J$. We consider the nested-surface case in (e) and (f) where we plot, respectively, the steady-state density distribution in momentum space and the phase distribution in real space, blue (orange) tiles corresponding to a $3\pi/4$ ($-\pi/4$) phase. The parameters are chosen as $\Delta=\gamma=0.1J$, $g=0$, and, $L=64$ (note that real-space plots zoom into the central area of the simulated domain, for which we chose periodic boundaries).
} \label{fig:fig2}
\end{figure*}

In order to determine the steady-state configuration of the system, we have numerically evolved Eqs. (\ref{eq:GP}) until they settle into some final state that we denote by $\lim_{t\rightarrow\infty}\psi_\mathbf{i}(t)\equiv\bar{\psi}_\mathbf{i}$.   We have exhaustively analyzed different random initial conditions, especially initial configurations randomly distributed around a uniform complex background $\psi_0$, that is, $\psi_\mathbf{i}(0)=\psi_0+\delta\psi_\mathbf{i}$, with $\delta\psi_\mathbf{i}$ having random phases and magnitudes uniformly distributed in the interval $[0,0.1|\psi_0|]$. For the parameters of interest, we have found that the steady-state properties are independent of the initial state. Of course, patterns spontaneously break the system's translational invariance, and can therefore emerge in any of several equivalent configurations (e.g., the orientation of the stripes), randomly selected by the initial fluctuations.

Note that in momentum space, the nonlinear terms induce scattering between different $\mathbf{k}$-modes, leading to a nonlinear competition that is won by modes located at the Bose surface, where the divergence rates are maximized. The geometry of such Bose surface plays then a crucial role in determining the spatial pattern the bosons condense to. In the following, we study two different Bose surfaces, depicted in Figs. \ref{fig:fig2}a and \ref{fig:fig2}d. We focus the numerics on moderate values of the interactions ($g<\gamma$, in particular), since otherwise the term $g|\psi_\mathbf{i}|^2$ might induce a shift of the chemical potential, and bring us off the Bose-surface geometry we are interested in. This regime is also aligned with realistic experimental conditions \cite{Leghtas15,Leghtas20,Wang20} in the implementation we propose below.

{\it Generic Bose surface versus Bose-surface nesting}.---We first consider the $\nu\neq 0$ case, for which the Bose surface forms a closed ring with $C_4$ rotational symmetry, see Fig. \ref{fig:fig2}a. Since the divergence rates of all the modes at the Bose surface are identical, one might expect a uniform density distribution of them. This is additionally supported by the fact that momentum conservation allows now for the so-called ``BCS'' scattering channel \cite{Shankar1994} $(\mathbf{k}_1,-\mathbf{k}_1)\rightarrow (\mathbf{k}_2,-\mathbf{k}_2)$, that couples arbitrary momenta $\pm\mathbf{k}_1$ and $\pm\mathbf{k}_2$ on the Bose surface.

This intuition is however challenged by our numerical results. We show in Fig. \ref{fig:fig2}b the steady-state density $n_\mathbf{k}=|\bar{\psi}_\mathbf{k}|^2$. In contrast to the expected uniform distribution on the Bose surface, a pair of $\pm\mathbf{k}$-modes is spontaneously selected by the random initial conditions as evidenced by the sharp peaks on the plot. In Fig. \ref{fig:fig2}c we show the corresponding real-space density {\clb $n_\mathbf{i}=|\bar{\psi}_\mathbf{i}|^2$}, which shows the corresponding striped pattern. In addition to the exhaustive numerical analysis, we have been able to prove analytically \cite{SupMat} that this striped patterns are stable against perturbations with momenta at the Bose surface, and also against small-momentum excursions, see below. In contrast, we prove \cite{SupMat} that even though the expected uniform solution exists, it is unstable.  Moreover, in \cite{SupMat} we show that the selected amplitudes have the fixed-phase relation $\bar{\psi}_{\mathbf{k}}=-i\bar{\psi}_{-\mathbf{k}}^*e^{-i\varphi}=e^{i\phi} \sqrt{\rho/3}$, where $\varphi=\text{arg}\{\gamma+ig\}$ and $\rho=L^2\Delta/\sqrt{\gamma^2+g^2}$. $\phi$ is an arbitrary phase that determines the location of the pattern, which is random ought to the translational invariance of the problem.

The most interesting situation occurs for the square lattice model with $\nu=0$, where the Bose surface contours coincide when shifted along a fixed reciprocal lattice vector $\mathbf{G}/2=(\pi,\pi)$, see Fig. \ref{fig:fig2}d. This effect, dubbed ``Fermi surface nesting'', is known to play an important role in determining the properties of the Fermi-Hubbard model at half-filling \cite{Hirsch1985}. One of the most important consequences of such effect is that the number of scattering channels increases dramatically, e.g., given three momenta on the Bose surface, one can always find a fourth one such that $\mathbf{k}_1+\mathbf{k}_2=\mathbf{k}_3+\mathbf{k}_4+\mathbf{G}$ (Umklapp scattering), see Fig. \ref{fig:fig2}d. Such scattering channels are allowed in the lattice system since the total momentum is shifted by a reciprocal lattice vector during the scattering process. In the closed fermionic model these new channels are responsible for the gap opening and the divergence of the density wave susceptibility at momentum $\mathbf{G}/2$ \cite{Shankar1994}. Here, we show that they can also significantly change the properties of the non-equilibrium steady state of our bosonic model.

\begin{figure}[t]
\includegraphics[width=0.99\linewidth]{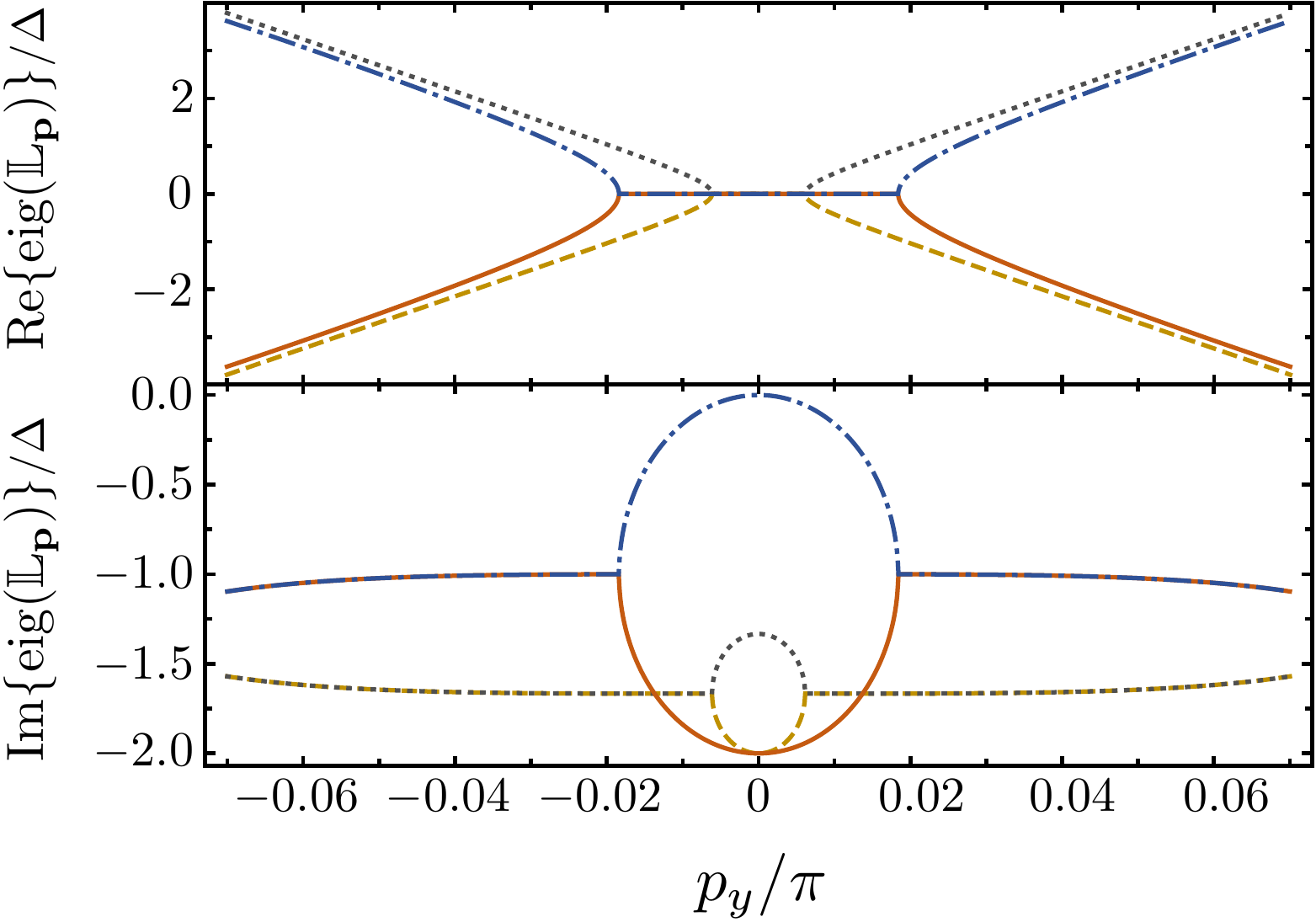}
\caption{Real (top) and imaginary (bottom) parts of the relaxation spectrum as a function of momentum excursions $p_y$ (for $p_x=0$). We consider a square lattice with $\nu=-J$, $\Delta=0.1J$, $g=0$, and $\mathbf{k}=(0,2\pi/3)$. A single imaginary eigenvalue (blue, dashed-dotted line) dominates the spectrum around $\mathbf{p}=\mathbf{0}$.}
\label{fig:fig3}
\end{figure}

The steady-state density distribution $n_\mathbf{k}$ is plotted in Fig. \ref{fig:fig3}e, where we see that, in contrast to the previous generic Bose surface where condensation occurs only on two $\pm\mathbf{k}$-modes, here all the modes on the Bose surface are occupied. Such a steady state is an unconventional BEC, with bosons condensed on a closed ring, instead of discrete points. In turn, the real-space density distribution $n_\mathbf{i}$ is completely uniform  \cite{SupMat}, while the phase distribution $\phi_\mathbf{i}=\arg\{\bar{\psi}_\mathbf{i}\}$ follows the rule that each lattice site must have two pairs of neighbors differing by a $\pi$ phase, which creates nontrivial phase portraits (Fig. \ref{fig:fig2}f). We have been able to derive this solution analytically, even proving that it is robust against arbitrary perturbations \cite{SupMat}. In equilibrium physics, bosons usually prefer to condense into discrete points to avoid exchange energy. Only under very specific conditions (e.g., moat-like band structures with infinitely-degenerate minima forming a closed curve), it is  conjectured that the interplay between the degeneracy and quantum correlations leads to a Bose-liquid state of the type we have found here \cite{Sedrakyan2015,Jiang2019}. In our non-equilibrium case, such unconventional superfluid states have a completely different origin: a momentum selection mechanism induced by the interplay between non-equilibrium conditions, nonlinear mode couplings, and lattice effects. Energy minimization is no longer criterion here since the steady state in our model is not related to any ground state.

{\clb Given the qualitative difference between the steady states for generic and nested Bose surfaces, one may wonder how they are connected as $\nu$ approaches zero. In the parameter regime that we study, $g<\gamma\ll J$, the interaction-induced shift of $\nu$ can be neglected, and thus the physics is dominated by its bare value. As a consequence, the transition between these two steady states occurs suddenly at $\nu=0$ within the mean-field approximation. For different parameter regimes (e.g., the strongly interacting case $g\sim J$), the shift in $\nu$ and corrections to the mean-field theory must become relevant, leading to a more complicated transition. Possible scenarios include that in which the discontinuous transition is turned into a crossover with coexistence of both states, or that in which the original transition point at $\nu=0$ is extended into a stable intermediate phase where the $\pm\mathbf{k}$ peaks of the stripped pattern continuously broaden as $\nu$ is reduced towards 0.}

{\it Relaxation spectrum}.---It is interesting to understand the way in which perturbations relax towards the steady-state condensate. To this aim, and as shown in detail in \cite{SupMat}, we transform Eq. (\ref{eq:GP}) to momentum space, and linearize it with respect to fluctuations around a generic Bose surface where bosons have condensed into a pair of modes with opposite momenta $\pm\mathbf{k}_0$. Specifically, we expand the amplitudes as
\begin{equation}
\psi_\mathbf{k}(t)=\bar{\psi}_{\mathbf{k}_0}\delta_{\mathbf{k}_0\mathbf{k}}+\bar{\psi}_{-\mathbf{k}_0}\delta_{-\mathbf{k}_0\mathbf{k}}+d_\mathbf{k}(t)	,
\end{equation}
and consider only fluctuations with small-momentum excursions $\mathbf{p}$ around $\pm\mathbf{k}_0$, that is, $|\mathbf{p}|\ll|\mathbf{k}_0|$. This leads to a closed linear system $i\dot{\mathbf{d}}_\mathbf{p}=\mathbb{L}_\mathbf{p}\mathbf{d}_\mathbf{p}$ for the fluctuations $\mathbf{d}_\mathbf{p}=(d_{\mathbf{k}_0+\mathbf{p}},d_{-\mathbf{k}_0+\mathbf{p}},d_{\mathbf{k}_0-\mathbf{p}}^*,d_{-\mathbf{k}_0-\mathbf{p}}^*)^T$, with a relaxation matrix $\mathbb{L}_\mathbf{p}$ that we provide in \cite{SupMat}. The eigenvalues of this matrix determine the relaxation spectrum, and are plotted in Fig. \ref{fig:fig3} for one characteristic example. For all choice of parameters we find that relaxation is dominated by a single eigenvalue, which can be approximated by a purely-imaginary quadratic form $-i\mathbf{p}^T\mathcal{K}\mathbf{p}$. The curvature matrix $\mathcal{K}$ depends on the system parameters, but the result is otherwise universal, indicating a purely diffusive, non-propagating behavior of the elementary excitations of our open system, similarly to what has been shown for exciton-polariton condensates \cite{Littlewood2006,Wouters2007}. By exhaustive inspection we have found that the striped patters are stable (i.e., $\mathcal{K}$ has positive eigenvalues) for $g<\gamma$, but can be destabilized when $g>\gamma$, leading to more complicated patterns, which we will study in the future. We have also checked that our results are robust against linear dissipation as long as nonlinear dissipation dominates.

{\it Experimental implementation.}---We propose to implement our model with an array of superconducting circuits known as transmons \cite{Krantz19}, which act as weakly-nonlinear quantum oscillators, discussed in more detail in \cite{SupMat}. Pair driving and dissipation are well established for these circuits \cite{Leghtas15,Leghtas20,Wang20}, where we remark that the ``chemical potential'' $\nu$ becomes easily tunable through external fields \cite{SupMat}. In addition, current chips allow for 2D lattices with as many as 54 transmons and tunable couplings, as demonstrated in Google's pioneering experiments leading to quantum advantage \cite{Supremacy19}. This number keeps growing steadily motivated by the goal of practical quantum computing. Moreover, we remark that transmon arrays have already allowed for proof-of-principle experiments exploring the standard BH model in 1D \cite{Schuster19}.

We emphasize that our work reveals the intriguing possibility that quantum computation platforms are not only of immense practical significance, but also pose their own interest as analog quantum simulators of emergent many-body phenomena far from equilibrium.

{\it Discussion}.---We comment now on the relation and differences between our results and other relevant work. Stripe phases, as a consequence of condensation on a pair of modes with opposite momenta, have been observed in both equilibrium \cite{lin2011} and non-equilibrium \cite{Clark2016} closed interacting bosonic systems. In both cases, the momenta correspond to the energy minimum of an effective Hamiltonian ({\it e.g.} a Floquet Hamiltonian for periodically driven systems \cite{Clark2016}). In contrast, in our driven-dissipative model, the pair of momenta is spontaneously selected among extensive degenerate modes at the Bose surface, which is formed by the maximally-divergent momenta, and thus has nothing to do with the minimum of any Hamiltonian. Hexagonal patterns \cite{Ardizzone2013} and solitons \cite{Ma17} have been observed in continuous-space driven-dissipative exciton-polaritons \cite{Carusotto2013}, in this case emerging from the interplay between linear losses, interactions, and a judicious spatio-temporal choice of driving fields. In our system, nonlinear dissipation and the lattice effect are crucial for the stabilization of exotic states with bosons condensed on a closed ring. This state is of great relevance for some open problems in condensed matter, and has not been predicted before by any other driven-dissipative mechanism to our knowledge. Currently, lattices can be engineered on exciton-polariton systems \cite{Kim2011,Jacqmin14,StJean20}, opening the possibility of implementing our ideas on such platforms as well.

{\it Conclusions and outlook}.---In this work we have studied the steady states of a pair-driven-dissipative BH model of relevance for current quantum simulators based on superconducting-circuit arrays, and leading to unconventional superfluid states of relevance for condensed-matter.  We have shown that the shape of a so-called ``Bose surface'' is crucial for the stead-state properties of driven-dissipative bosonic systems, reminiscing the behavior of interacting fermions at equilibrium. Future developments will include the analysis of models with flat bands (i.e., bands with constant $\varepsilon_\mathbf{k}$), where bosons can potentially condense into spatially-localized structures such as solitons.

\begin{acknowledgements}
{\it Acknowledgments}.---We thank Germ\'an J. de Valc\'arcel for useful suggestions. ZC is supported in part by the National Key Research and Development Program of China (Grant No. 2016YFA0302001), NSFC of  China (Grant No. 11674221, No.11574200), the Project of Thousand Youth Talents, the Program Professor of Special Appointment (Eastern Scholar) at Shanghai Institutions of Higher Learning and the Shanghai Rising-Star program. We also acknowledge additional support from a Shanghai talent program and Shanghai Municipal Science and Technology Major Project (Grant No.2019SHZDZX01)
\end{acknowledgements}

%

\newpage

\begin{widetext}

\begin{center}
\textbf{\Large{Supplemental material}}
\end{center}

In this supplemental material we provide a more detailed view of the proposed experimental implementation, including the master equation that describes it, and how it leads to the equations that we have used in the main text under the coherent-state approximation for the condensate. Next we analytically study the three types of steady-state condensate solutions that we have mentioned in the text: trivial, density waves, and uniform on the Bose surface, including their stability.

\vspace{3mm}
\begin{center}
\textbf{\large{}I. From the master equation in the laboratory frame to our model equations}{\large\par}
\end{center}

As mentioned in the main text, for the implementation of our ideas we consider an array of superconducting circuits called ``transmons'' \cite{Krantz19,Supremacy19,Schuster19}, which we sketch and describe in Fig. \ref{fig:Circuits}. The Josephson junctions present in the transmon circuits makes them behave as coupled nonlinear oscillators, whose dynamics is described by the Hamiltonian
\begin{equation}
{\clb\frac{\hat{H}_\text{transmons}}{\hbar}}=\sum_\mathbf{i}\left(\omega_0\hat{b}_\mathbf{i}^\dagger\hat{b}_\mathbf{i}+\frac g2 \hat{b}_\mathbf{i}^{\dagger 2}\hat{b}_\mathbf{i}^2\right)-J\sum_{\langle \mathbf{ij}\rangle}(\hat{b}^\dag_\mathbf{i} \hat{b}_\mathbf{j}+\text{H.c.}),
\end{equation}
where $\omega_0\gg(g,J)$ is the bare frequency of the transmons and $g$ is the nonlinear coefficient induced by the junction. In addition, we consider each transmon to be coupled to an external transmission line (that doesn't host frequency $\omega_0$), which is strongly driven by two coherent tones at frequencies $\omega_1<\omega_0$ and $\omega_2>\omega_0$ such that $\omega_1+\omega_2=2(\omega_0-\nu)$. $\nu$ is then the detuning of the four-wave mixing process that takes two excitations from the drives (one from each) and turns them into two excitations of the transmon, or vice versa \cite{Leghtas15,Leghtas20,Wang20}. This parameter will play the role of the chemical potential of our model, which can hence be tuned experimentally at will and at real time, simply by detuning appropriately the coherent tones.
\begin{center}
\begin{figure}[b]
\includegraphics[width=0.4\textwidth]{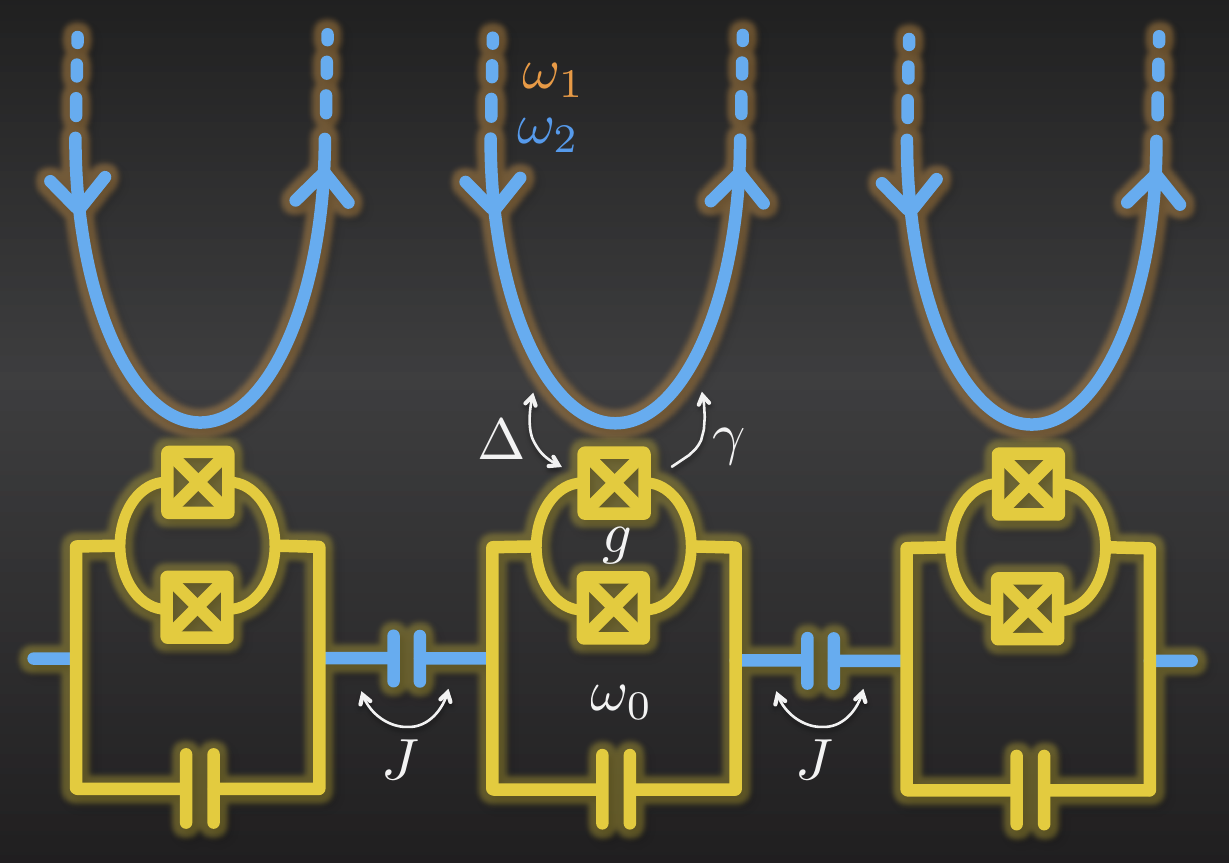}
\caption{Schematic representation of the proposed experimental implementation. The basic elements are transmons (in yellow), which are superconducting circuits formed by Josephson junctions (in a squid configuration in the figure) and capacitors, and behave as quantum oscillators with bare frequency $\omega_0$ and weak anharmonicity $g$ \cite{Krantz19}. The transmons are capacitively coupled, inducing a tunneling rate $J$ (in experiments the coupling can be mediated by an auxiliary biased LC circuit, which makes $J$ tunable \cite{Krantz19,Supremacy19}). In addition, each transmon is coupled to a transmission line that allows the propagation of two coherent fields with frequencies $\omega_1<\omega_0$ and $\omega_2>\omega_0$ such that $\omega_1+\omega_2\approx 2\omega_0$, but which doesn't host frequency $\omega_0$ (in experiments, auxiliary LC circuits that act as filters are commonly employed for this matter \cite{Leghtas15,Leghtas20,Wang20}). The nonlinearity of the junctions induces a four-wave mixing process that allows the transmon excitations to decay in pairs into the line at rate $\gamma$, while the external fields exchange coherently pairs of excitations at rate $\Delta$ with the transmons. The effective chemical potential of the model is given by the detuning $\nu=\omega_0-(\omega_1+\omega_2)/2$.}\label{fig:Circuits}
\end{figure}
\end{center}

\vspace{-8mm}
The transmission lines can be treated as an environment, which can be formally integrated, leading to a model for the transmons alone. In particular, the common approach in quantum optics consists on starting from the von Neumann equation $i{\clb\hbar} d\hat{\rho}_\text{total}/dt=[\hat{H}_\text{total},\hat{\rho}_\text{total}]$, where $\hat{\rho}_\text{total}$ and $\hat{H}_\text{total}$ are, respectively, the state and Hamiltonian describing the total system (transmons + transmission lines), including their interaction; then, under the usual Born-Markov approximation, the environment is traced out, obtaining a so-called ``master equation'' for the trasnmon's state alone $\hat{\rho}_\text{transmons}=\text{tr}_\text{lines}\{\hat{\rho}_\text{total}\}$. In our case, this leads to
\begin{equation}
\frac{d\hat{\rho}_\text{transmons}}{dt}=-i\left[{\clb\frac{\hat{H}_\text{transmons}}{\hbar}}+\sum_\mathbf{i}\frac{\Delta}{2}\left(e^{i(\omega_1+\omega_2)t}\hat{b}_\mathbf{i}^2+e^{-i(\omega_1+\omega_2)t}\hat{b}_\mathbf{i}^{\dagger 2}\right),\hat{\rho}_\text{transmons}\right]+\frac{\gamma}{2}\sum_\mathbf{i}\mathcal{D}_{b_\mathbf{i}^2}[\hat{\rho}_\text{transmons}],
\end{equation}
where we have defined the Lindblad form $\mathcal{D}_J[\hat{\rho}]=2\hat{J}\hat{\rho}\hat{J}^\dagger-\hat{J}^\dagger\hat{J}\hat{\rho}-\hat{\rho}\hat{J}^\dagger\hat{J}$. The parameter $\gamma$ is proportional to the coupling between the transmission lines and the transmons, while the parameter $\Delta$ is proportional the amplitude of the driving fields, and can therefore be controlled experimentally at real time. Note that this means that all the parameters of the model can be experimentally adjusted independently.

The master equation above is explicitly time dependent. However, moving to picture rotating at the driving frequency $(\omega_1+\omega_2)/2$, we obtain an autonomous problem. In particular, the state $\hat{\rho}=\hat{U}^\dagger\hat{\rho}_\text{transmons}\hat{U}$, with $\hat{U}=\exp[-i(\omega_1+\omega_2)t\sum_\mathbf{i}\hat{b}_\mathbf{i}^\dagger\hat{b}_\mathbf{i}/2]$, evolves according to the master equation
\begin{equation}
\frac{d\hat{\rho}}{dt}=-i\Bigg[\underset{\hat{H}{\clb/\hbar}}{\underbrace{\sum_\mathbf{i}\left(\frac g2 \hat{b}_\mathbf{i}^{\dagger 2}\hat{b}_\mathbf{i}^2-\nu\hat{b}_\mathbf{i}^\dagger\hat{b}_\mathbf{i}+\frac{\Delta}{2}\left(\hat{b}_\mathbf{i}^2+\hat{b}_\mathbf{i}^{\dagger 2}\right)\right)-J\sum_{\langle \mathbf{ij}\rangle}(\hat{b}^\dag_\mathbf{i} \hat{b}_\mathbf{j}+\text{H.c.})}},\hat{\rho}\Bigg]+\frac{\gamma}{2}\sum_\mathbf{i}\mathcal{D}_{b_\mathbf{i}^2}[\hat{\rho}].
\end{equation}
Note that this master equation can be written in the alternative form
\begin{equation}
\frac{d\hat{\rho}}{dt}=-{\clb\frac{i}{\hbar}}\left(\hat{H}_\text{eff}\hat{\rho}-\hat{\rho}\hat{H}_\text{eff}^\dagger\right)+\gamma\sum_\mathbf{i}\hat{b}_\mathbf{i}^2\hat{\rho}\hat{b}_\mathbf{i}^{\dagger 2},\label{EffMaster}
\end{equation}
where
\begin{equation}
\hat{H}_\text{eff}=\hat{H}-i\frac{{\clb\hbar}\gamma}{2}\sum_\mathbf{i}\hat{b}_\mathbf{i}^{\dagger 2}\hat{b}_\mathbf{i}^2,	
\end{equation}
can be interpreted as an effective non-Hermitian Hamiltonian. The last term in Eq. (\ref{EffMaster}) accounts for irreversible ``quantum jumps'', which are required in order to preserve the normalization of the state.

Whenever these jumps are negligible or play no role, the description of the open system based on a non-Hermitian Hamiltonian is reasonable. This is the case, for example, when the bosons form a Bose-Einstein condensate. The reason for this is that the state of the condensate is approximately coherent, $|\Psi\rangle=\prod_\mathbf{i}\exp(\psi_\mathbf{i}\hat{b}_\mathbf{i}^\dagger-\psi_\mathbf{i}^*\hat{b}_\mathbf{i})|0\rangle$, which is an eigenstate of the annihilation operators, $\hat{b}_\mathbf{i}|\Psi\rangle=\psi_\mathbf{i}|\Psi\rangle$, so that quantum jumps have no effect on it ($|0\rangle$ is the vacuum state). This is the approach that we adopted in the main text. In particular, in order to find the condensate's steady-state configuration we have made a coherent-state ansatz with time-dependent amplitudes $\psi_\mathbf{i}(t)$, whose evolution equation can be found as follows. First, note that evolution equation of the expectation value of any operator $\hat{B}$ can be written as
\begin{equation}
\frac{d}{dt}	\langle\hat{B}\rangle=\mathrm{tr}\left\{\hat{B}\frac{d\hat{\rho}}{dt}\right\}=-i\left\langle\left[\hat{B},\frac{\hat{H}}{\hbar}\right]\right\rangle+\frac{\gamma}{2}\sum_\mathbf{j}\left(\left\langle\left[\hat{b}_\mathbf{j}^{\dagger 2},\hat{B}\right]\hat{b}_\mathbf{j}^2\right\rangle+\left\langle\hat{b}_\mathbf{j}^{\dagger 2}\left[\hat{B},\hat{b}_\mathbf{j}^2\right]\right\rangle\right).
\end{equation}
Applying it to the annihilation operators $\hat{b}_\mathbf{i}$, and assuming that the state is coherent at all times, so that $\hat{b}_\mathbf{i}|\Psi\rangle=\psi_\mathbf{i}|\Psi\rangle$ and $\langle\Psi|\hat{b}_\mathbf{i}^\dagger=\langle\Psi|\psi_\mathbf{i}^*$, we obtain the GP-like evolution equations presented in the main text, which we reproduce here for convenience:
\begin{equation}
\frac{d\psi_\mathbf{i}}{dt}=iJ\sum_{\mathbf{j}\in\langle\mathbf{i}\rangle}\psi_\mathbf{j}+i\nu\psi_\mathbf{i}-(\gamma+ig)|\psi_\mathbf{i}|^2\psi_\mathbf{i}-i\Delta \psi_\mathbf{i}^*, \label{GPeqSup}
\end{equation}
where we have introduced the notation $\langle\mathbf{i}\rangle$ for the sites adjacent to $\mathbf{i}$.

As mentioned in the main text, this equation describes very well the dynamics of the system as long as the system is in a superfluid state and in the thermodynamic limit of infinite number of bosons. In our model, superfluid order is expected to appear when the hopping rate is the dominant scale (which we have assumed throughout the main text), as experimentally demonstrated in exciton-polariton platforms \cite{Carusotto2013,Ardizzone2013,Ma17} and theoretically discussed in \cite{Wouters2007,Tauber2014}. Note that for a finite-size system, master equations usually have a unique mixed steady state. However, in systems with spontaneous continuous-symmetry breaking, one also finds infinitely-many metastable states, whose dissipative tunneling rate and decay rate into the true steady state decreases with the size of the system \cite{Kinsler91,Navarrete17cycles,Iemini18}. Hence, the thermodynamic limit brings infinitely-many symmetry-breaking ordered steady states, which are stable against symmetry-breaking perturbations. {\clb It is in this limit where the pure coherent-state ansatz is expected to describe the physics of the superfluid phase correctly.}

\vspace{3mm}
\begin{center}
\textbf{\large{}II. GP equations in reciprocal space: steady-state and stability equations}{\large\par}
\end{center}

In order to analyze the different kinds of stationary solutions that the equations above have and their stability, it is convenient to  transform them to reciprocal space. Using the relations (we define for convenience the number of lattice sites $N=L^2$)
\begin{equation}
\psi_{\mathbf{k}}=\frac{1}{\sqrt{N}}\sum_{\mathbf{i}}e^{-i\mathbf{k}\cdot\mathbf{i}}\psi_{\mathbf{i}}\hspace{2mm}\Leftrightarrow\hspace{2mm}\psi_{\mathbf{i}}=\frac{1}{\sqrt{N}}\sum_{\mathbf{k}}e^{i\mathbf{k}\cdot\mathbf{i}}\psi_{\mathbf{k}},\label{GPsup}	
\end{equation}
between the amplitudes in real and reciprocal space, and the identity $\sum_{\mathbf{i}}e^{i\mathbf{k}\cdot\mathbf{i}}=N\delta_{\mathbf{k},\mathbf{0}}$, Eqs. (\ref{GPsup}) are turned into
\begin{equation}
\dot{\psi}_{\mathbf{k}}=-i(\varepsilon_{\mathbf{k}}-\nu)\psi_{\mathbf{k}}-i\Delta\psi_{-\mathbf{k}}^{*}-\frac{\gamma+ig}{N}\sum_{\mathbf{k}_{1}\mathbf{k}_{2}}\psi_{\mathbf{k}_{1}}\psi_{\mathbf{k}_{2}}\psi_{\mathbf{k}_{1}+\mathbf{k}_{2}-\mathbf{k}}^{*},\label{GPk}
\end{equation}
where we have defined the dispersion relation $\varepsilon_\mathbf{k}=-2J(\cos k_x+\cos k_y)$ on the square lattice. We will study the stationary solutions of these equations, $\lim_{t\rightarrow\infty}\psi_\mathbf{k}(t)\equiv\bar{\psi}_\mathbf{k}$.

The stability of any such stationary solution $\bar{\psi}_\mathbf{k}$ can be studied by analyzing the evolution of fluctuations around it. Expanding the modal amplitudes in Eq. (\ref{GPk}) as $\psi_\mathbf{k}(t)=\bar{\psi}_\mathbf{k}+d_\mathbf{k}(t)$, and keeping terms up to first order in the fluctuations $d_\mathbf{k}(t)$, we obtain the linear system
\begin{equation}
\dot{d}_{\mathbf{k}}=-i(\varepsilon_{\mathbf{k}}-\nu)d_{\mathbf{k}}-i\Delta d_{-\mathbf{k}}^{*}-\frac{\gamma+ig}{N}\sum_{\mathbf{k}_{1}\mathbf{k}_{2}}(2\bar{\psi}_{\mathbf{k}_{2}}\bar{\psi}_{\mathbf{k}_{1}+\mathbf{k}_{2}-\mathbf{k}}^{*}d_{\mathbf{k}_{1}}+\bar{\psi}_{\mathbf{k}_{1}}\bar{\psi}_{\mathbf{k}_{2}}d_{\mathbf{k}_{1}+\mathbf{k}_{2}-\mathbf{k}}^{*}).\label{GenStabilitySup}
\end{equation}
Whenever all fluctuations $d_\mathbf{k}(t)$ decay in time, the stationary solution $\bar{\psi}_\mathbf{k}$ is said to be stable.

\vspace{3mm}
\begin{center}
\textbf{\large{}III. Trivial solution and its stability}{\large\par}
\end{center}

The simplest stationary solution we can consider is the trivial one $\bar{\psi}_\mathbf{k}=0$. For this solution, the terms under the sum in Eq. (\ref{GenStabilitySup}) vanish, so that the stability is then completely set by the quadratic part of the Hamiltonian, which we have characterized in the main text. In particular, we showed that the trivial solution becomes unstable whenever there exist modes for which $\Delta>|\varepsilon_\mathbf{k}-\nu|$. This is indeed the case for the situations we consider in this work, since we assume that there are modes at the Bose surface ($\varepsilon_\mathbf{k}-\nu=0$), so that any $\Delta\neq 0$ will induce an instability, no matter how small.

\vspace{3mm}
\begin{center}
\textbf{\large{}III. Striped patterns}{\large\par}
\end{center}

\vspace{1mm}
\begin{center}
\textbf{III.A. Striped density-wave solutions}
\end{center}

As the simplest nontrivial solution, and motivated by our numerical findings, we consider the case in which the nonlinear competition is won by a density wave with underlying wave vectors $\pm\mathbf{k}_0$ at the Bose surface, that is,
\begin{equation}
\bar{\psi}_\mathbf{k}=\bar{\psi}_{\mathbf{k}_0}\delta_{\mathbf{k}_0\mathbf{k}}+\bar{\psi}_{-\mathbf{k}_0}\delta_{-\mathbf{k}_0\mathbf{k}}.\label{StripedSols}
\end{equation}
These are the type of solutions that have emerged numerically in the case of a generic Bose surface, with $\mathbf{k}_0$ spontaneously chosen from the available momenta at the Bose surface by the random initial fluctuations.

Using this ansatz, Eqs. (\ref{GPk}) are turned into the following pair of coupled equations for the density-wave amplitudes $\bar{\psi}_{\mathbf{k}_0}$ and $\bar{\psi}_{-\mathbf{k}_0}^*$:
\begin{subequations}
\begin{align}
0 & =\bar{\psi}_{-\mathbf{k}_{0}}^{*}-i\frac{\gamma+ig}{\Delta N}\left(2|\bar{\psi}_{-\mathbf{k}_{0}}|^{2}+|\bar{\psi}_{\mathbf{k}_{0}}|^{2}\right)\bar{\psi}_{\mathbf{k}_{0}},
\\
0 & =\bar{\psi}_{\mathbf{k}_{0}}+i\frac{\gamma-ig}{\Delta N}\left(2|\psi_{\mathbf{k}_{0}}|^{2}+|\psi_{-\mathbf{k}_{0}}|^{2}\right)\bar{\psi}_{-\mathbf{k}_{0}}^{*}.
\end{align}
\end{subequations}
Decomposing the parameter $(\gamma+ig)/\Delta N\equiv \alpha e^{i\varphi}$ in magnitude $\alpha=\sqrt{\gamma^2+g^2}/\Delta N$ and phase $\varphi=\arg\{\gamma+ig\}$, and similarly for the density-wave amplitudes, $\bar{\psi}_{\pm\mathbf{k}_0}=\rho_\pm e^{i\phi_\pm}$, the equations are turned into
\begin{subequations}
\begin{align}
\rho_- e^{-i(\phi_++\phi_-)}& =ie^{i\varphi}\alpha(2\rho_-^{2}+\rho_+^{2})\rho_+,
\\
\rho_+ e^{-i(\phi_++\phi_-)} & =-ie^{-i\varphi}\alpha(2\rho_+^{2}+\rho_-^2)\rho_-.
\end{align}
\end{subequations}
Taking absolute values, we are left with two coupled equations for the magnitudes $\rho_\pm$, with only one nontrivial, real, and positive solution: $\rho_\pm=1/\sqrt{3\alpha}$. On the other hand, the equations above only fix the phase sum $\phi_-+\phi_+=-\varphi-\pi/2$, with the phase difference remaining arbitrary. This allows us to write the final solution as we did in the main text:
\begin{equation}
\bar{\psi}_{\mathbf{k}_{0}}=\sqrt{\frac{1}{3\alpha}}e^{i\phi},\hspace{3mm}\bar{\psi}_{-\mathbf{k}_{0}}=-i\bar{\psi}_{\mathbf{k}_{0}}^{*}e^{-i\varphi},\label{StripeSols2}	
\end{equation}
where $\phi$ is arbitrary. Note that the corresponding density in real space reads $n_\mathbf{i}\sim\cos^2(\mathbf{k}_0\cdot\mathbf{i}+\phi+\varphi/2+\pi/4)$, so that different choices of $\phi$ lead to the same pattern of stripes forming an angle $\arctan(k_{0x}/k_{0y})$ with respect to the $x$ axis, but with maxima shifted to different positions. The choice of $\phi$ by the random initial fluctuations provides then an example of spontaneous symmetry breaking of spatial translations.

\vspace{1mm}
\begin{center}
\textbf{III.B. Stability of the striped patterns and relaxation equations}
\end{center}

We can analyze the stability of these striped density waves by particularizing Eqs. (\ref{GenStabilitySup}) to the solution of Eqs. (\ref{StripedSols}) and (\ref{StripeSols2}). Noting that in such case
\begin{subequations}
\begin{align}
\sum_{\mathbf{k}_{1}\mathbf{k}_{2}}\bar{\psi}_{\mathbf{k}_{2}}\bar{\psi}_{\mathbf{k}_{1}+\mathbf{k}_{2}-\mathbf{k}}^{*}d_{\mathbf{k}_{1}} & =2|\bar{\psi}_{\mathbf{k}_{0}}|^{2}d_{\mathbf{k}}+\bar{\psi}_{\mathbf{k}_{0}}\bar{\psi}_{-\mathbf{k}_{0}}^{*}d_{\mathbf{k}-2\mathbf{k}_{0}}+\bar{\psi}_{\mathbf{k}_{0}}^{*}\bar{\psi}_{-\mathbf{k}_{0}}d_{\mathbf{k}+2\mathbf{k}_{0}},\\
\sum_{\mathbf{k}_{1}\mathbf{k}_{2}}\bar{\psi}_{\mathbf{k}_{1}}\bar{\psi}_{\mathbf{k}_{2}}d_{\mathbf{k}_{1}+\mathbf{k}_{2}-\mathbf{k}}^{*} & =\bar{\psi}_{\mathbf{k}_{0}}^{2}d_{2\mathbf{k}_{0}-\mathbf{k}}^{*}+2\bar{\psi}_{-\mathbf{k}_{0}}\bar{\psi}_{\mathbf{k}_{0}}d_{-\mathbf{k}}^{*}+\bar{\psi}_{-\mathbf{k}_{0}}^{2}d_{-2\mathbf{k}_{0}-\mathbf{k}}^{*},
\end{align}	
\end{subequations}
we then find
\begin{equation}
\dot{d}_{\mathbf{k}}=i\left(\varepsilon_{\mathbf{k}}-\nu-\frac{4\Delta}{3}e^{i\varphi}\right)d_{\mathbf{k}}+\frac{i\Delta}{3}\left[-d_{-\mathbf{k}}^{*}+2e^{2i(\varphi+\phi)}d_{\mathbf{k}-2\mathbf{k}_{0}}+2e^{-2i\phi}d_{\mathbf{k}+2\mathbf{k}_{0}}+e^{i(\varphi+2\phi)}d_{2\mathbf{k}_{0}-\mathbf{k}}^{*}+e^{-i(\varphi+2\phi)}d_{-2\mathbf{k}_{0}-\mathbf{k}}^{*}\right].\small\label{GenStabilityK0}
\end{equation}
While it is not easy to find a closed form for the eigenvalues of this linear system, we can do so by considering two specific types of perturbations.

\begin{center}
\textbf{Perturbations at the Bose surface}
\end{center}

As a first case of stability analysis that we can treat analytically, we consider perturbations with momenta $\pm\mathbf{k}\neq\pm\mathbf{k}_0$ on the Bose surface. For this, we simply particularize (\ref{GenStabilityK0}) to those modes (for which $\varepsilon_{\mathbf{k}}-\nu=0$), taking into account that $\mathbf{k}\pm 2\mathbf{k}_{0}$ are not at the Bose surface in the generic case, so that $d_{\mathbf{k}\pm 2\mathbf{k}_{0}}=0=d_{-\mathbf{k}\pm 2\mathbf{k}_{0}}$. This leads to a simple linear system $\dot{\mathbf{d}}=\mathbb{L}\mathbf{d}$ for $\mathbf{d}=(d_\mathbf{k},d_{-\mathbf{k}}^*)$, with stability matrix
\begin{equation}
\mathbb{L}=-\frac{\Delta}{3}\left(\begin{array}{cc}
4e^{i\varphi} & -i\\
i & 4e^{-i\varphi}
\end{array}\right),
\end{equation}
whose eigenvalues have negative real part for any choice of $\varphi\in[0,\pi/2]$, showing that the striped patterns are stable against perturbations with momenta at the Bose surface.

\newpage
\begin{center}
\textbf{Small-momentum excursions}
\end{center}

As a second example that allows for a semi-analytic treatment, we consider fluctuations that perform only small-momentum excursions around $\pm\mathbf{k}_{0}$. Specifically, particularizing Eq. (\ref{GenStabilityK0}) to $\mathbf{k}=\mathbf{k}_{0}+\mathbf{p}$, we obtain
\begin{align}
\dot{d}_{\mathbf{k}_{0}+\mathbf{p}}&=-i\left(\varepsilon_{\mathbf{k}_{0}+\mathbf{p}}-\nu-\frac{4\Delta}{3}e^{i\varphi}\right)d_{\mathbf{k}_{0}+\mathbf{p}}+\frac{i\Delta}{3}\left(-d_{-\mathbf{k}_{0}-\mathbf{p}}^{*}+2e^{2i(\varphi+\phi)}d_{-\mathbf{k}_{0}+\mathbf{p}}+e^{i(\varphi+2\phi)}d_{\mathbf{k}_{0}-\mathbf{p}}^{*}\right)
\\
&\hspace{5.6cm}+\frac{i\Delta}{3}\left(2e^{-2i\phi}d_{3\mathbf{k}_{0}+\mathbf{p}}+e^{-i(\varphi+2\phi)}d_{-3\mathbf{k}_{0}-\mathbf{p}}^{*}\right).\nonumber
\end{align}
Assuming $|\mathbf{p}|\ll\mathbf{k}_{0}$ and considering only fluctuations around $\pm\mathbf{k}_{0}$, we can drop the terms in the second line. Proceeding in the same way for the other three possibilities $\mathbf{k}_0-\mathbf{p}$ and $-\mathbf{k}_0\pm\mathbf{p}$, we then find the closed linear system $i\dot{\mathbf{d}}_{\mathbf{p}}=\mathbb{L}_{\mathbf{p}}\mathbf{d}_{\mathbf{p}}$
for the fluctuations $\mathbf{d}_{\mathbf{p}}=(d_{\mathbf{k}_{0}+\mathbf{p}},d_{-\mathbf{k}_{0}+\mathbf{p}},d_{\mathbf{k}_{0}-\mathbf{p}}^{*},d_{-\mathbf{k}_{0}-\mathbf{p}}^{*})^T$, with
\begin{equation}
\mathbb{L}_{\mathbf{p}}=\left(\begin{array}{cccc}
\varepsilon_{\mathbf{k}_{0}+\mathbf{p}}-\nu-\frac{4i\Delta}{3}e^{i\varphi} & \frac{2\Delta}{3}e^{2i\varphi} & -\frac{i\Delta}{3}e^{i\varphi} & \frac{\Delta}{3}
\\
-\frac{2\Delta}{3} & \varepsilon_{-\mathbf{k}_{0}+p}-\nu-\frac{4i\Delta}{3}e^{i\varphi} & \frac{\Delta}{3} & \frac{i\Delta}{3}e^{-i\varphi}
\\
-\frac{i\Delta}{3}e^{-i\varphi} & -\frac{\Delta}{3} & -\varepsilon_{\mathbf{k}_{0}-\mathbf{p}}+\nu-\frac{4i\Delta}{3}e^{-i\varphi} & -\frac{2\Delta}{3}e^{-2i\varphi}
\\
-\frac{\Delta}{3} & \frac{i\Delta}{3}e^{i\varphi} & \frac{2\Delta}{3} & -\varepsilon_{-\mathbf{k}_{0}-\mathbf{p}}+\nu-\frac{4i\Delta}{3}e^{-i\varphi}
\end{array}\right),	
\end{equation}
where we have set $\phi=0$ since the eigenvalues do not depend on this arbitrary phase (in fact, it can be absorbed in the fluctuations $d_\mathbf{k}$). The imaginary part of the eigenvalues of this matrix determines the stability of the striped patterns. As a function of $\mathbf{p}$, this is sometimes called the ``spectrum of elementary excitations'' of the open system, although a more precise term that we advocate for would be ``relaxation spectrum''. While the explicit form of the eigenvalues is too large to print it here, they are easily found analytic, which has allowed us to study them exhaustively. As explained in the main text, we have found that the relaxation spectrum is dominated by a single eigenvalue that can be approximated by a quadratic form $-i\mathbf{p}^T\mathcal{K}\mathbf{p}$ on the momentum excursions $\mathbf{p}$, where the $2\times 2$ curvature matrix $\mathcal{K}$ depends on the system parameters $\nu/\Delta$, $J/\Delta$, $\mathbf{k}_0$, and $\varphi$. Our exhaustive numerical analysis has concluded that for moderate values of the transmon's nonlinearity $g<\gamma$ (corresponding to $\varphi<\pi/4$), the eigenvalues of $\mathcal{K}$ are positive, signaling that the stripe patterns are stable. This is no longer true when $g>\gamma$ ($\varphi>\pi/4$), for which even small momentum excursions along certain directions can grow towards other types of patterns. It will be interesting to understand in the future which kind of patterns can be generated this way.

\vspace{3mm}
\begin{center}
\textbf{\large{}IV. Uniform solutions on a generic Bose surface}{\large\par}
\end{center}

Another interesting family of solutions is that in which all the modes of the Bose surface are equally populated. As explained in the text, this seems a natural solution as well, because all of them have the same divergence rate $\Delta$. We then consider stationary solutions of the type
\begin{equation}
\bar{\psi}_{\mathbf{k}}=\left\{\begin{array}{cc}
\rho e^{i\phi_{\mathbf{k}}} & \text{for }\mathbf{k}\in\text{Bose surface}
\\
0 & \text{for }\mathbf{k}\notin\text{Bose surface}
\end{array}\right.,\label{UniformSol}
\end{equation}
such that the density $n_\mathbf{k}=|\psi_\mathbf{k}|^2=\rho^2$ at the Bose surface is uniform. Noting that for such solution we have
\begin{equation}
\sum_{\mathbf{k}_{1}\mathbf{k}_{2}}\bar{\psi}_{\mathbf{k}_{1}}\bar{\psi}_{\mathbf{k}_{2}}\bar{\psi}_{\mathbf{k}_{1}+\mathbf{k}_{2}-\mathbf{k}}^{*} =\left[2\left(\sum_{\mathbf{q}}|\bar{\psi}_{\mathbf{q}}|^{2}\right)-|\bar{\psi}_{\mathbf{k}}|^{2}\right]\bar{\psi}_{\mathbf{k}}+\left(\sum_{\mathbf{q}\neq\pm\mathbf{k}}\bar{\psi}_{\mathbf{q}}\bar{\psi}_{-\mathbf{q}}\right)\bar{\psi}_{-\mathbf{k}}^{*},\label{NonlinearSum}
\end{equation}
where the final sums run only over modes at the Bose surface (as do the ones in what follows), we can plug the ansatz (\ref{UniformSol}) in Eq. (\ref{GPk}), obtaining
\begin{equation}
1-\alpha\rho^{2}\sum_{\mathbf{q}\neq\pm\mathbf{k}}e^{i(\phi_{\mathbf{q}}+\phi_{-\mathbf{q}}+\varphi+\pi/2)}=\alpha\rho^{2}(2N_{\text{BS}}-1)e^{i(\phi_{\mathbf{k}}+\phi_{-\mathbf{k}}+\varphi+\pi/2)},
\end{equation}
where $N_{\text{BS}}=\sum_{\mathbf{q}}\propto\sqrt{N}$ is the number of modes on the Bose surface. While this equation could have solutions with complicated phase profiles $\phi_\mathbf{k}$, the most natural solution is obtained by assuming $\phi_{\mathbf{k}}+\phi_{-\mathbf{k}}=-\varphi-\pi/2$ $\forall\mathbf{k}$, as happened with the striped patterns. With that assumption, and using $\sum_{\mathbf{q}\neq\pm\mathbf{k}}=-2+\sum_{\mathbf{q}}=N_{\text{BS}}-2$, we obtain the density
\begin{equation}
\rho^{2}=\frac{1}{3\alpha(N_{\text{BS}}-1)}.\label{UniformDensity}	
\end{equation}
Dividing the Bose surface into the upper and lower halves, denoted by $\text{BS}_{\pm}$, respectively, the final solution reads
\begin{equation}
\bar{\psi}_{\mathbf{k}}=\left\{ \begin{array}{cl}
\rho e^{i\bar{\phi}_{\mathbf{k}}} & \text{for }\mathbf{k\in}\text{BS}_{+}\\
\rho e^{-i(\bar{\phi}_{\mathbf{k}}+\varphi+\pi/2)} & \text{for }\mathbf{k\in}\text{BS}_{-}\\
0 & \text{for }\mathbf{k\notin}\text{BS}
\end{array}\right.,\label{UniformBSsol}
\end{equation}
where the phases $\bar{\phi}_{\mathbf{k}}$ are arbitrary, and we denote by BS the whole Bose surface. This leads to a very complex density in real space
\begin{equation}
n_{\mathbf{i}}\sim\left[\sum_{\mathbf{k}\in\text{BS}_{+}}\cos\left(\mathbf{k}\cdot\mathbf{i}+\bar{\phi}_{\mathbf{k}}+\frac{\varphi}{2}+\frac{\pi}{4}\right)\right]^{2},
\end{equation}
where now the arbitrary phases $\bar{\phi}_{\mathbf{k}}$ play a crucial role in determining not only on the location of the pattern, but also the shape of the pattern itself.

Even though this solution exists as a fixed point and seems like a natural one, we have been able to prove analytically that it is unstable. For this, we consider a very specific type of perturbations that we describe next. First, we consider perturbations only along the Bose surface, that is, $d_{\mathbf{k}\notin\text{BS}}=0$. In addition, we consider perturbations that preserve the phase relations of the stationary solution, that is, $d_{-\mathbf{k}}=-id_{\mathbf{k}}^{*}e^{-i\varphi}$. Next, we consider only density fluctuations, that is, $d_{\mathbf{k}\in\text{BS}_+}(t)=r_{\mathbf{k}}(t)e^{i\bar{\phi}_{\mathbf{k}}}$,
with $r_{\mathbf{k}}\in\mathbb{R}$ and $\bar{\phi}_{\mathbf{k}}$ the phase
of the stationary solution (\ref{UniformBSsol}). Note that this means that the amplitudes of the upper half of the Bose surface read
\begin{equation}
\psi_{\mathbf{k}}(t)=[\rho+r_{\mathbf{k}}(t)]e^{i\bar{\phi}_{\mathbf{k}}}.\label{PsiDensityFluc}
\end{equation}
Finally, we consider density fluctuations $r_{\mathbf{k}}(t)$ that create an imbalance between the population of one specific pair of
modes $\pm\mathbf{k}_{0}$ and the rest of the Bose surface, but leaving the total density at the Bose surface invariant. This is accomplished by giving them the form
\begin{equation}
r_{\mathbf{k}}(t)=\left(\delta_{\mathbf{k}\mathbf{k_{0}}}-\frac{1-\delta_{\mathbf{k}\mathbf{k_{0}}}}{N_{\text{BS}}/2-1}\right)r(t),\label{DensityFluctuations}
\end{equation}
indeed note that the total Bose-surface density $n_\text{BS}(t)=\sum_{\mathbf{k}\in\text{BS}}|\psi_{\mathbf{k}}(t)|^{2}$ is not affected by the local density fluctuations (up to first order in $r$):
\begin{equation}
n_\text{BS}=2\sum_{\mathbf{k}\in\text{BS}_{+}}|\psi_{\mathbf{k}}|^{2}\approx N_\text{BS}\rho^{2}+4\rho\sum_{\mathbf{k}\in\text{BS}_{+}}r_{\mathbf{k}}=N_{\text{BS}}\rho^{2}+4\rho r\left(1-\frac{1}{N_{\text{BS}}/2-1}\sum_{\mathbf{k}\in\text{BS}_{+}\neq\mathbf{k}_{0}}\right)=N_{\text{BS}}\rho^{2},
\end{equation}
where we have used $\sum_{\mathbf{k}\in\text{BS}_{+}\neq\mathbf{k}_{0}}=-1+\sum_{\mathbf{k}\in\text{BS}_{+}}=N_{\text{BS}}/2-1$. In order to determine the evolution equation of $r(t)$, the fastest route is to go back to the GP equations (\ref{GPk}), particularized to the case in which only modes at the Bose surface are populated, that is, $\psi_{\mathbf{k}\notin\text{BS}}=0$. Using Eq. (\ref{NonlinearSum}) and the relation $\psi_{-\mathbf{k}}=-i\psi_{\mathbf{k}}^{*}e^{-i\varphi}$, so that
\begin{equation}
ie^{i\varphi}\sum_{\mathbf{q}\in\text{BS}\neq\pm\mathbf{k}_{0}}\psi_{\mathbf{q}}\psi_{-\mathbf{q}}=\sum_{\mathbf{q}\in\text{BS}\neq\pm\mathbf{k}_{0}}|\psi_{\mathbf{q}}|^{2}=-2|\psi_{\mathbf{k}}|^{2}+\sum_{\mathbf{q}\in\text{BS}}|\psi_{\mathbf{q}}|^{2},	
\end{equation}
the GP equation takes the simpler form
\begin{equation}
\dot{\psi}_{\mathbf{k}}=\Delta e^{i\varphi}\left(1+3\alpha|\psi_{\mathbf{k}}|^{2}-6\alpha\sum_{\mathbf{q}\in\text{BS}_{+}}|\psi_{\mathbf{q}}|^{2}\right)\psi_{\mathbf{k}},
\end{equation}
which we have already written in terms of modes at the upper half of the Bose surface only. Inserting (\ref{PsiDensityFluc}) and (\ref{DensityFluctuations}) in this expression,
 we finally obtain a simple evolution equation for the density fluctuations $r(t)$:
\begin{equation}
 \dot{r}=6\alpha\rho^{2}\Delta e^{i\varphi}r\approx\frac{2\Delta}{N_{\text{BS}}}e^{i\varphi}r,	
\end{equation}
which provides a growth rate $6\alpha\rho^{2}\Delta\cos\varphi>0$ (remember that $\varphi\in[0,\pi/2]$),
showing that the stationary solution we have considered is unstable.

\vspace{3mm}
\begin{center}
\textbf{\large{}V. Spatially-uniform solutions for Bose-nested surfaces}{\large\par}
\end{center}

In the main text we have shown that the phenomenon of Bose surface nesting appears for a square lattice with $\nu=0$, opening up a massive amount of scattering channels that end up populating all the modes of the Bose surface. For such case, we have shown how the numerics lead to a uniform density in real space, $n_\mathbf{i}=\bar{n}$ $\forall\mathbf{i}$. Here we analytically find such solution and prove that it is stable. We also prove that this solution is not available for $\nu\neq 0$.

Let us then start by assuming a solution of the type $\bar{\psi}_{\mathbf{i}}=\sqrt{\bar{n}} e^{i\phi_{\mathbf{i}}}$. Introducing this ansatz into Eq. (\ref{GPeqSup}), we get
\begin{equation}
J\sum_{\mathbf{j}\in\langle\mathbf{i}\rangle}e^{i\phi_{\mathbf{j}}}=\Delta e^{-i\phi_{\mathbf{i}}}-i(\gamma+ig)\bar{n}e^{i\phi_{\mathbf{i}}},	
\end{equation}
where we use again the notation $\langle\mathbf{i}\rangle$ for the sites adjacent to $\mathbf{i}$. As we prove at the end of the section, the condition that only modes at the Bose surface are populated is equivalent to the demand $J\sum_{\mathbf{j}\in\langle\mathbf{i}\rangle}e^{i\phi_\mathbf{j}}=0$. We then assume this to hold, and later check that the solution we find is consistent with it. Under such assumption, we then find two solutions to the equation above
\begin{equation}
\bar{\psi}_{\mathbf{i}}=\pm\sqrt{\bar{n}} e^{-i(\varphi/2+\pi/4)},\hspace{3mm}\text{with }\bar{n}=\frac{\Delta}{\sqrt{\gamma^2+g^2}},\label{HomSol}	
\end{equation}
where we remind that $\varphi=\arg\{\gamma+ig\}$. Note that each site $\mathbf{i}$ can choose between the `$+$' or `$-$' solutions independently. Hence, the spatially-uniform solution is indeed compatible with the assumption we made above, as long as each lattice site has the same number of $+$ and $-$ neighbors. For example, a simple phase profile with all sites at even (odd) rows choosing the $+$ ($-$) solution satisfies this. However, starting from random initial conditions, more intricate phase profiles are found, such as the one showed in the main text.

Next let's check the stability of the solution. Expanding the real-space amplitudes around the stationary solution as $\psi_{\mathbf{i}}(t)=\bar{\psi}_{\mathbf{i}}+d_{\mathbf{i}}(t)$, the GP equation (\ref{GPeqSup}) leads to the following equation to first order in the perturbations $d_\mathbf{i}$:
\begin{equation}
i\dot{d}_{\mathbf{i}}=-J\sum_{\mathbf{j}\in\langle\mathbf{i}\rangle}d_{\mathbf{j}}+\Delta d_{\mathbf{i}}^{*}-i(\gamma+ig)\left(2|\bar{\psi}_{\mathbf{i}}|^{2}d_{\mathbf{i}}+\bar{\psi}_{\mathbf{i}}^{2}d_{\mathbf{i}}^{*}\right),	
\end{equation}
which for the solution we are analyzing reads
\begin{equation}
\dot{d}_{\mathbf{i}}=i\left[J\sum_{\mathbf{j}\in\langle\mathbf{i}\rangle}d_{\mathbf{j}}-2\Delta\sin(\varphi)d_\mathbf{i}\right]-2\Delta\cos(\varphi)d_{\mathbf{i}}.	
\end{equation}
This provides a closed set of linear equations for the perturbations $d_\mathbf{i}$, where all have the same decay rate $2\Delta\cos\varphi>0$ (remember once again that $\varphi\in[0,\pi/2]$). Therefore, we conclude that the spatially-uniform solution is stable against arbitrary perturbations.

It is important to remark that this solution is not available for the case of a generic Bose surface, that is, for $\nu\neq 0$. This is because in such case, the condition that only modes at the Bose surface are populated is equivalent to $J\sum_{\mathbf{j}\in\langle\mathbf{i}\rangle}e^{i\phi_\mathbf{j}}=-\nu e^{i\phi_\mathbf{i}}$, as we prove next, which is at odds with solution (\ref{HomSol}). Let us show how this condition comes about. First, we note that Fourier transform relates the first two terms of Eq. (\ref{GPeqSup}) (GP equation in real space) with the first term of Eq. (\ref{GPsup}) (GP equation in momentum space), that is
\begin{equation}
-J\sum_{\mathbf{j}\in\langle\mathbf{i}\rangle}\psi_{\mathbf{j}}-\nu\psi_{\mathbf{i}}=\frac{1}{\sqrt{N}}\sum_{\mathbf{k}}(\varepsilon_{\mathbf{k}}-\nu)\psi_{\mathbf{k}}e^{i\mathbf{k}\cdot\mathbf{i}}.\label{FT}	
\end{equation}
On the other hand, note that the stationary solution in momentum space and the dispersion relation satisfy
\begin{equation}
\bar{\psi}_{\mathbf{k}}=\left\{ \begin{array}{cc}
\neq0 & \mathbf{k}\in\text{BS} \\ =0 & \mathbf{k}\notin\text{BS}
\end{array}\right.,\hspace{1em}\text{and}\hspace{1em}\varepsilon_{\mathbf{k}}-\nu=\left\{ \begin{array}{cc}
=0 & \mathbf{k}\in\text{BS}\\
\neq0 & \mathbf{k}\notin\text{BS}
\end{array}\right.,	
\end{equation}
so that the right-hand-side of Eq. (\ref{FT}) vanishes when particularized to the steady-state solution, leading to
\begin{equation}
J\sum_{\mathbf{j}\in\langle\mathbf{i}\rangle}\bar{\psi}_{\mathbf{j}}=-\nu\bar{\psi}_{\mathbf{i}},	
\end{equation}
which provides the expression we wanted to prove considering that $\bar{\psi}_{\mathbf{i}}=\sqrt{\bar{n}} e^{i\phi_{\mathbf{i}}}$.

\newpage
\end{widetext}
\end{document}